\documentclass{article}

\usepackage{spconf,amsmath,graphicx}

\usepackage{enumitem}
\setlist{nosep, leftmargin=14pt}

\usepackage{mwe} 
\usepackage{subcaption}
\usepackage{hyperref}

\title{A Unified Approach for Comprehensive Analysis of Various Spectral and Tissue Doppler Echocardiography}
\name{
{Jaeik Jeon$^{1 \star}$ \thanks{$\star$ denotes the first author that contributed equally to this work, while $\dagger$ denotes the corresponding author (jyg1722@gmail.com; yeonyeeyoon@gmail.com).} \qquad 
Jiyeon Kim$^{2,3 \star}$ \qquad 
Yeonggul Jang$^{1 \dagger}$ \qquad 
Yeonyee E. Yoon $^{4 \dagger}$} \qquad  
Dawun Jeong $^{2,3}$   \\
{{\it Youngtaek Hong$^{1,3}$ } \qquad 
{\it Seung-Ah Lee $^1$} \qquad 
{\it Hyuk-Jae Chang$^{1,3,5}$}}
}

\address{$^{1}$ Ontact Health, Seoul, South Korea  \\
    $^{2}$Department of Internal Medicine, Graduate School of Medical Science,\\ Brain Korea 21 Project, Yonsei University College of Medicine\\ 
    $^{3}$ CONNECT-AI Research Center, Yonsei University College of Medicine, Seoul, South Korea\\
    $^{4}$Cardiovascular Center, Seoul National University Bundang Hospital, Seongnam, South Korea \\
    $^{5}$Severance Cardiovascular Hospital, Yonsei University Health System, Seoul, South Korea
} 

\begin{document}
%
\maketitle
\begin{abstract}
Doppler echocardiography offers critical insights into cardiac function and phases by quantifying blood flow velocities and evaluating myocardial motion. However, previous methods for automating Doppler analysis, ranging from initial signal processing techniques to advanced deep learning approaches, have been constrained by their reliance on electrocardiogram (ECG) data and their inability to process Doppler views collectively. We introduce a novel unified framework using a convolutional neural network for comprehensive analysis of spectral and tissue Doppler echocardiography images that combines automatic measurements and end-diastole (ED) detection into a singular method. The network automatically recognizes key features across various Doppler views, with novel Doppler shape embedding and anti-aliasing modules enhancing interpretation and ensuring consistent analysis. Empirical results indicate a consistent outperformance in performance metrics, including dice similarity coefficients (DSC) and intersection over union (IoU). The proposed framework demonstrates strong agreement with clinicians in Doppler automatic measurements and competitive performance in ED detection.
\end{abstract}
\section{Introduction}
\label{sec:intro}
Doppler echocardiography is pivotal in assessing cardiac function, particularly through its ability to capture dynamic, time-dependent changes in velocity. Spectral Doppler effectively maps the velocity and direction of blood flow over time, while Tissue Doppler Imaging (TDI) is adept at measuring the time-variant velocity of myocardial tissue. These modalities, through the analysis of spectral and tissue Doppler imaging, provide critical clinical metrics, including maximum blood flow velocity ($V_{max}$) and velocity time integral (VTI). Importantly, the full spectrum of data these techniques offer extends well beyond these commonly measured indicators, capturing a comprehensive temporal dynamics of cardiac cycles.

\begin{figure*}[ht!]
\centering
\normalsize
\centerline{\includegraphics[width=1.8\columnwidth]{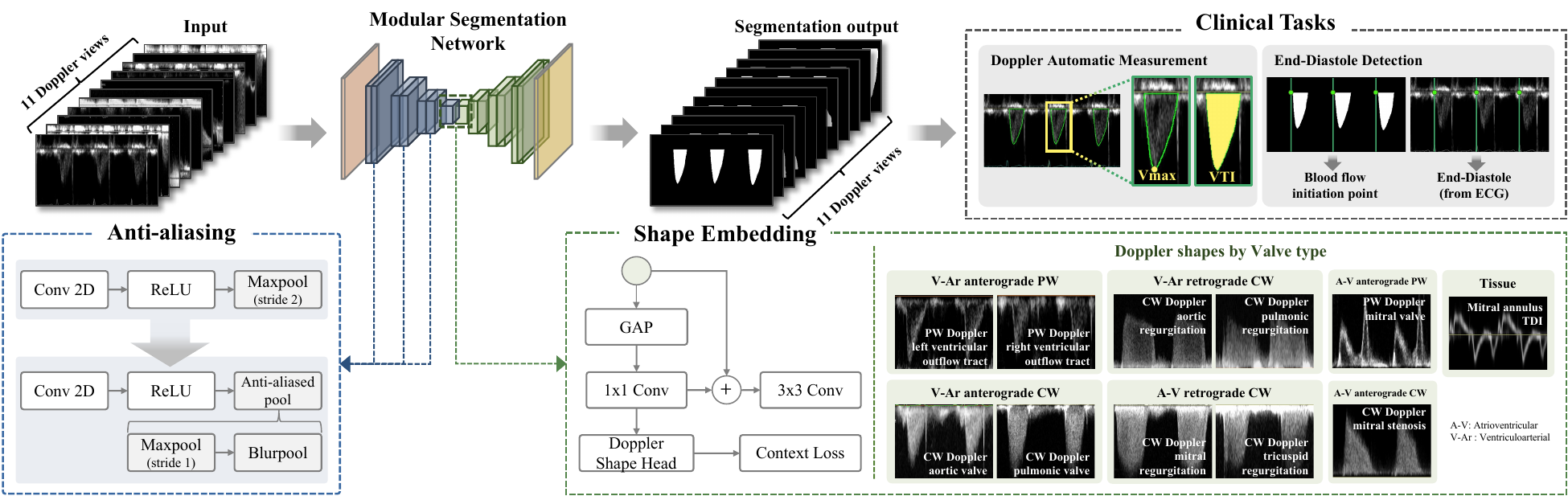}}
\captionof{figure}{{\bf{Unified framework for comprehensive analysis}}. 
The network, with its anti-aliasing and Doppler shape embedding modules, accurately processes diverse Doppler views and supports clinical tasks like automatic measurement and ED detection.
}
\label{fig 1}
\end{figure*}

Since early works, efforts to automatically measure clinical parameters from Doppler images have been made. These approaches \cite{park2008automatic, zolgharni2014automated,  taebi2018estimating} primarily relied on digital and signal processing, involving steps such as noise filtering to obtain the Doppler envelope and thresholding to detect key points for obtaining clinical measurements. Nonetheless, these algorithms were often compromised by poor contrast and image artifacts, and required hyperparameter tuning for each views, making them less effective and creating significant challenges for comprehensive automated Doppler analysis.

Recent deep learning-based methods have advanced the analysis of Doppler images, notably improving the classification of Doppler types \cite{sulas2018automatic, gilbert2020user}, the evaluation of Doppler flow quality \cite{zamzmi2019echo}, and the automation of mitral inflow velocity measurements \cite{elwazir2020fully}. Nevertheless, these methods depend on electrocardiograms (ECG) for determining cardiac phases or identifying regions of interest, and are restricted to processing each Doppler view individually. The continued dependence on ECG and the view-specific limitation indicate only modest progress. Despite efforts to automatically detect end-diastole (ED) \cite{jahren2020estimation}, full integration with automated measurement is lacking. This gap highlights the pressing need for a unified framework that would bring together all aspects of Doppler echocardiography analysis. Such a framework would not only facilitate cardiac phase recognition and extensive automatic measurement across the full spectrum of Doppler views but also achieve this without the need for ECG or other auxiliary inputs.

In this work, we present a novel unified framework that enables comprehensive analysis of various spectral and tissue Doppler echocardiography images using a single fully convolutional network (Fig. \ref{fig 1}). To equip the network to discern the temporal dynamics of cardiac cycles without ECG, we have trained the segmentation network using annotated segmentation masks that mimic the VTI. This segmentation is pivotal; it not only captures key topological features for clinical measurement but also integrates information on the cardiac phase, given that VTI is characterized as the integral of the velocity curve over a cardiac cycle.  Our single network is trained on a comprehensive dataset that covers the full spectrum of Doppler modalities, including both pulsed wave (PW) and continuous wave (CW) Doppler, as well as extending to TDI, thus enabling it to autonomously learn and discern key features across these varied views. To effectively support an integrated interpretation of diverse Doppler signal data, we propose a Doppler shape embedding module. Additionally, we propose the integration of an anti-aliasing module to ensure baseline-shift equivariance, which is essential for maintaining consistent analysis despite variations in baseline positioning. Our tailored modules demonstrate consistent performance increases compared to networks without these enhancements in metrics such as Dice similarity coefficients (DSC) and Intersection over union (IoU). Our comprehensive approach not only excels in automatic measurements and ED detection but also sets a new standard by eliminating the need for ECG data, a significant step beyond previous ECG-dependent and single-view methods.

\section{Methods}
\label{sec:Methods}
In the section, we provide a detailed description of the segmentation network, which incorporates the Doppler shape embedding module alongside an anti-aliasing strategy. We will then outline the methodology for extracting clinical parameters, such as $V_{max}$ and VTI. Moreover, we will present a clinical application: a method for detecting ED.

\subsection{Doppler envelope segmentation}
 The segmentation architecture is reinforced by two key modules: a Doppler Shape Embedding module and an anti-aliasing module, each tailored to overcome distinct challenges in interpreting Doppler signals.

The Doppler shape embedding module, inspired by insights from prior research \cite{yu2021bisenet}, is designed to capture the shape features intrinsic to Doppler spectrograms for each Doppler view. As depicted in Fig. \ref{fig 1}, Doppler signals can be classified into seven distinct flow types based on valve positions and the direction of blood flow – whether anterograde or retrograde. Accordingly, the shape embedding block is strategically designed to incorporate this contextual knowledge, particularly at the final stage of the encoder, possessing high-level semantic details. This module, situated at the encoder's final stage, applies global average pooling followed by a $1\times1$ convolution, enhancing the semantic features related to Doppler shapes. This integration of contextual knowledge, along with a shape head for signal pattern classification and correction via context loss, ensures precise feature weighting corresponding to the diverse Doppler signals.

The anti-aliasing module tackles the issue of baseline shift. Clinicians often adjusts the Doppler signals’s vertical position to better visualize regions of interest (e.g., dominant flow), as shown in Fig. \ref{fig 2}. However, a recent study \cite{zou2023delving} reports that small input translations or rescaling significantly affect modern network’s prediction. By incorporating an anti-aliasing strategy, our network maintains robust performance, producing consistent segmentations even when the Doppler signal is manually adjusted, thus optimizing the utility of echocardiographic analysis in clinical practice. In this paper, we use Blurpool, proposed in \cite{zhang2019making}, as our anti-aliasing strategy. Blurpool consists of two operations: blurring filter with kernel k × k and subsampling. Therefore, we replace every max-pooling and a strided convolution operation with Blurpool, enabling the segmentation model to be anti-aliased and have a consistent output for baseline shift. 

\noindent
\begin{minipage}{1\columnwidth}
\begin{center}
\includegraphics[width=1\textwidth]{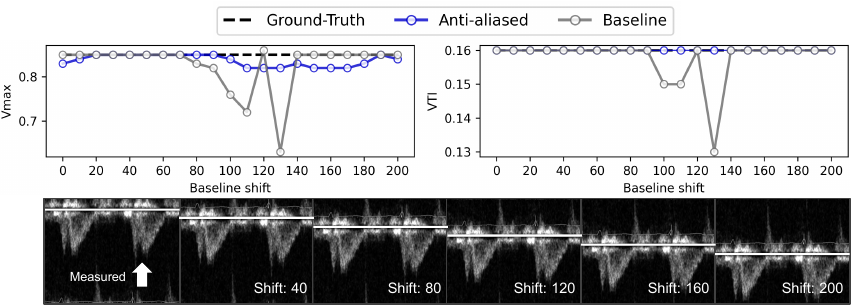}
\captionof{figure}{{ \bf{Performance comparison of the segmentation network with and without the anti-aliasing module, against baseline shifts.}}
}
\label{fig 2}
\end{center}

\end{minipage}

\

The total network optimization loss,  $L_{\textrm{total}}$, combines segmentation loss,  $L_{\textrm{seg}}$, and context loss,  $L_{\textrm{context}}$, weighted by a factor  $\mu$, and is defined as  $L_{\textrm{total}} = L_{\textrm{seg}} + \mu L_{\textrm{context}}$, with both losses calculated via cross-entropy.

\subsection{Modular Segmentation Networks}
Our framework's modularity facilitates the seamless integration of Doppler Shape Embedding and anti-aliasing modules into any existing encoder-decoder based convolutional architectures. To evaluate our approach, we engaged in comparative analyses with renowned segmentation networks, including Unet \cite{ronneberger2015u}, UNet++ \cite{zhou2018unet++}, and BiSeNetV2 \cite{yu2021bisenet}. 

\subsection{Clinical Parameters Computation and ED detection}
Given our segmentation network's training protocol, acquiring clinical parameters becomes remarkably straightforward. Once we obtain the segmentation mask, the VTI measurement naturally falls into place. This mask diligently pinpoints velocity profiles within our designated regions of interest, using the baseline for contextual reference. The maximum velocity is then measured by identifying the pixel positioned at the greatest distance from this baseline within the segmented Doppler signal. 

In our approach, the detection of ED is based on the observation of blood flow initiation or termination. For instance, in PW or CW Doppler of the atrioventricular (A-V) valves (mitral or tricuspid valve), the termination of anterograde flow indicates the closure of the A-V valve, delineating the ED, a detail that our Doppler envelope segmentation clearly capture. Similarly, the initiation of anterograde flow in PW or CW Doppler across ventriculoarterial (V-Ar) valves (aortic or pulmonic valve) marks the beginning of ventricular systole and simultaneously denotes the ED. Traditional methods, like those seen in \cite{jahren2020estimation}, determine these critical phases using the R-peak from ECG data. However, our method offers a more direct and potentially precise way to determine these events means of identifying these events by directly utilizing the timing of valve movements. A comprehensive breakdown of how the start or end of flow in specific views determines the ED can be found in Table \ref{table 2}.

\section{Experiments}
Our research employed a dataset of 25,854 Doppler DICOM files from 6,854 patients provided by the OpenAI Dataset Project (AI-Hub), a South Korean Ministry of Science and ICT initiative \cite{AI-hub}. In collaboration with clinicians, 11 clinically-relevant views were selected based on the ASE guidelines. Sonographers annotated segmentation masks for training and validating our network, which were also used to derive clinical parameters such as $V_{max}$ and VTI. ED labeling for ED detection performance evaluation was initially automated using the R-peak detection algorithm 
from ECG with subsequent manual verification. The data was split at the patient level into training (80\%), validation (10\%), and testing (10\%) sets. Employing the Monte Carlo cross-validation method, we repeated the dataset split five times to thoroughly evaluate our model's segmentation accuracy. For the segmentation performance, we report the average values derived from multiple dataset splits, and for the evaluation of Doppler measurements and end-diastole (ED) detection, we present the results from the first split of the test set.

\subsection{Implementation Details}
Our experiments were carried out using PyTorch. Doppler envelope is cropped from the original Dicom image using the header information (tag 0018, 6011), and resized to 256 × 512. Inputs were resized to the same value in the validation and test phase, preserving the original ratio. All image values were scaled between -1 and 1. No additional augmentations were used. The segmentation networks were trained using the Adam optimizer with learning rate of 1e-3. Both the proposed and comparison methods are trained for 200 epochs with a batch size of 32 and early stopped training when the validation segmentation loss stops improving. 

\subsection{Doppler Envelope segmentation}
\begin{figure}[t!]
  \centering
    \includegraphics[width=0.45\textwidth]{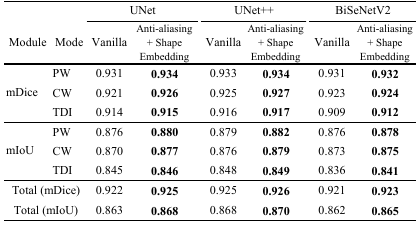}
\captionof{table}{{ \bf{Comparative performance analysis across three different modalities.}}}
\label{table 1}
\end{figure}

The effectiveness of our proposed modules on segmentation performance was assessed using DSC and IoU. Table \ref{table 1} displays the comparison between our enhanced models with proposed modules and baseline models, showing that our modules consistently outperforms network performance. The improvements are particularly noticeable in CW Doppler, with metrics like DSC and IoU consistently higher. This indicates that our Doppler shape embedding module, which integrates shape context, is key to extracting semantically rich features.

Furthermore, Fig. \ref{fig 2} shows that incorporating the anti-aliasing module  effectively counters baseline-shifts, maintaining automatic measurement performance. The absence of the anti-aliasing module results in significant discrepancies in both $V_{max}$ and VTI whenever baseline shifts occur. In contrast, the application of BlurPool demonstrates a notable reduction in deviations from the GT, yielding more consistent outputs. This outcome underscores the module's efficacy and validates its capacity to faithfully capture and reflect the inherent traits and variations present in Doppler images.

\noindent



\subsection{Clinical Tasks}

\subsubsection{Automatic Measurements} 
We employed Pearson correlation coefficients (PCC) to determine the agreement with clinicians. Ground truth (GT) segmentation masks serve as the basis for extracting these clinical measurements. The PCC was calculated by matching cardiac beats between the GT and the predictions, and the true detection rate ($\textrm{TDR}_{\textrm{measure}}$) was assessed by the ratio of number of correctly matched predictions to the GT count. 

In Table \ref{table 2}, a consistent and notably high correlation was observed for $V_{max}$ and VTI across all views. Fig. 3 (a) further illustrates this trend through a scatter plot, highlighting the outstanding agreement  between the model and clinician.

\subsubsection{End-diastole Detection} 
For ED detection performance, a detection limit parameter $\lambda$ was set , classifying predictions as accurate if the estimated ED fell within $\lambda$ of the R-peak in the ECG. The $\textrm{TDR}_{\textrm{ED}}$, defined as the ratio of correctly detected EDs to the entire count of labeled EDs, was then calculated, excluding the EDs within  100ms of boundaries, aligning with the evaluation presented in \cite{jahren2020estimation}.

We present results obtained with the parameter $\lambda$ at 0.08sec in Table \ref{table 2}. The data overall are quite promising. Particularly, examining PW or CW Doppler of A-V valves (mitral or tricuspid valve),  we observe $\textrm{TDR}_{\textrm{ED}}$ approaching 100\%. This aligns with the theoretical expectations for ED detection. On the other hand, we observe a decrease in the accuracy of ED detection in the context of CW Doppler of V-Ar valve (aortic or pulmonic valve) regurgitation. This can be attributed to the characteristic prolongation of regurgitant flow through the V-Ar valve, which extends beyond the ED phase and into the isovolumic contraction time. Fig. \ref{figure 3} (b) provides a graphical representation of $\textrm{TDR}_{\textrm{ED}}$ as a function of $\lambda$, offering insights into sensitivity. This visual aid helps us understand the variations in TDR based on different $\lambda$ values.

\

\noindent
\begin{minipage}{1\columnwidth}
\begin{center}
\includegraphics[width=1\textwidth]{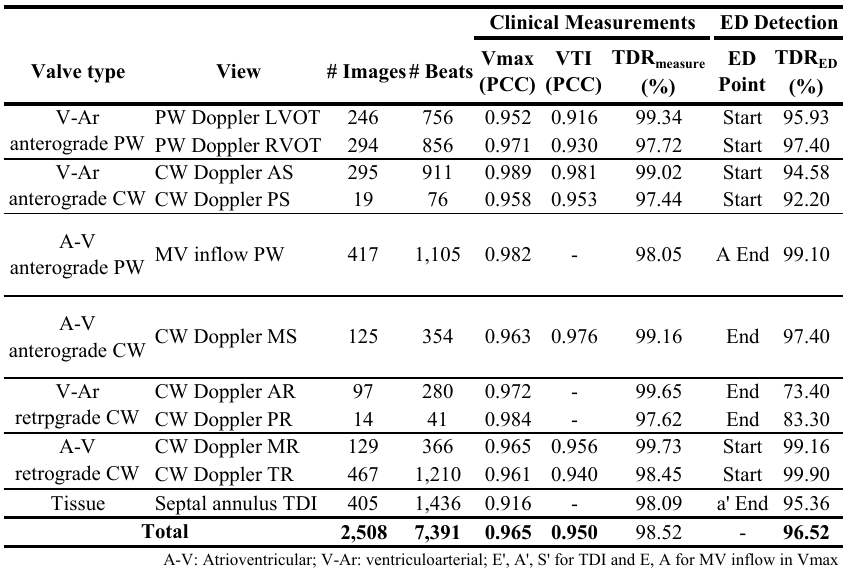}
\captionof{table}{{ \bf{Summary of automatic measurement correlations and end-diastole detection rates across various Doppler views.}}
}
\label{table 2}
\end{center}
\end{minipage}
\begin{minipage}{1\columnwidth}
\begin{center}
\includegraphics[width=1\textwidth]{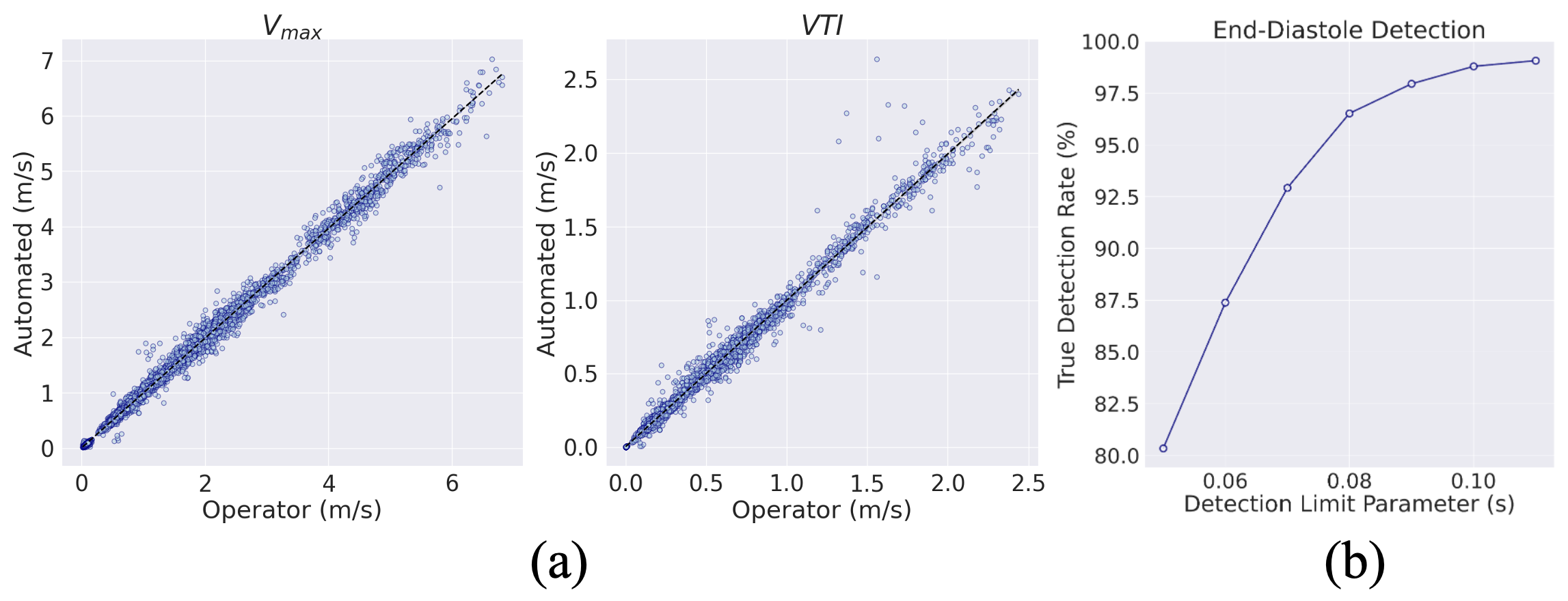}
\captionof{figure}{{ {
{\bf (a)} Scatter plots display strong correlations between $V_{max}$ and VTI predictions with clinician assessments. {\bf (b)} A curve shows the model's TDR for ED across different $\lambda$ thresholds.
}}
}
\label{figure 3}
\end{center}
\end{minipage}

\section{Conclusions}
Our study presents a unified framework for comprehensive analysis of various Doppler modalities, facilitating automatic measurements and ED detection without ECG reliance, though future work will aim to enhance its measurement capabilities and external validation.

\section{Compliance with ethical standards}
\label{sec:ethics}
This research study was conducted using the OpenAI Dataset Project (AI-Hub) collected from five tertiary hospitals, and the institutional review board of each hospital approved the use of de-identified data and waived the requirement for informed consent owing to the retrospective and observational study design (IRB No. 2021-0147- 003; CNUH 2021-04-032; HYUH 2021-03-026-003; SCHBC 2021-03-007-001; B-2104/677-004).


\section{Acknowledgments}
\label{sec:acknowledgments}

This work was supported by the Institute of Information \& communications Technology Planning \& Evaluation (IITP) grant funded by the Korean government (MSIT) (No.
\

\noindent
$2022000972$, development of flexible  mobile health care softwar eplatform using 5G MEC).



\bibliographystyle{IEEEbib}
\bibliography{strings,refs}

\end{document}